\documentclass[12pt,preprint]{aastex}
\shorttitle{Buoyancy-Driven Photon Bubbles}
\shortauthors{Begelman}
 
\shorttitle{Buoyancy-driven photon bubbles}

\shortauthors{Begelman}

\begin{document}
 
\title{Nonlinear Photon Bubbles Driven by Buoyancy}
\author{Mitchell C. Begelman\altaffilmark{1}}
\affil{Joint Institute for Laboratory Astrophysics, University of Colorado at Boulder}
\affil{JILA, 440 UCB, Boulder, CO 80309-0440} 
\email{mitch@jila.colorado.edu}
\altaffiltext{1}{Also at Department of Astrophysical and Planetary Sciences, University of Colorado}
       
 
\begin{abstract}
We derive an analytic model for nonlinear ``photon bubble" wave trains driven by buoyancy forces in magnetized, radiation pressure-dominated atmospheres.  Continuous, periodic wave solutions exist when radiative diffusion is slow compared to the dynamical timescale of the atmosphere.  We identify these waves with the saturation of a linear instability discovered by Arons --- therefore, these wave trains should develop spontaneously.  The buoyancy-driven waves are physically distinct from photon bubbles in the presence of rapid diffusion, which evolve into trains of gas pressure-dominated shocks as they become nonlinear.

Like the gas pressure-driven shock trains, buoyancy-driven photon bubbles can exhibit very large density contrasts, which greatly enhance the flow of radiation through the atmosphere.  However, steady-state solutions for buoyancy-driven photon bubbles exist only when an extra source of radiation is added to the energy equation, in the form of a flux divergence. We argue that this term is required to compensate for the radiation flux lost via the bubbles, which increases with height.  We speculate that an atmosphere subject to buoyancy-driven photon bubbles, but lacking this compensating energy source, would lose pressure support and collapse on a timescale much shorter than the radiative diffusion time in the equivalent homogeneous atmosphere.   
  
\end{abstract}

\keywords {accretion: accretion disks---hydrodynamics---instabilities---MHD---radiative transfer---X-rays: binaries}

\section{Introduction}

The tendency of ``photon bubbles" to form in magnetized, radiation pressure-supported atmospheres is well-established.  In photon bubble instability, the atmosphere spontaneously forms a propagating pattern of low-density channels separated by regions of high density.  Radiation tends to diffuse through the underdense regions, avoiding the regions of high density.  Both analytic calculations (Begelman 2001) and numerical simulations (Turner et al. 2005; Hsu, Arons \& Klein 1997) indicate that the instability can lead to very large and persistent density contrasts.  In this case, the atmosphere becomes ``leaky" and the net radiation flux can greatly exceed the equilibrium value for a homogeneous atmosphere.    

Two families of linear photon bubble instabilities were discovered by Arons (1992) and Gammie (1998; see also Blaes \& Socrates 2003), respectively.  The Gammie modes depend on the interplay of gas pressure and radiation pressure and are destabilized slow magnetosonic waves. In the nonlinear limit they become trains of gas pressure-dominated shocks.  The Arons modes are an outgrowth of internal entropy modes, and are related to previously-discovered instabilities driven by convection in a strong magnetic field, in the presence of heat conduction (Syrovatskii \& Zhugzhda 1968).  They are driven by buoyancy, are governed by the rate of radiative diffusion, and do not depend on the presence of gas pressure. The Arons modes occur where the parameter $M_0$, denoting the ratio of dynamical time to radiation diffusion time across the atmosphere, is smaller than one,  indicating that radiation is relatively well-coupled to the gas. In contrast, the Gammie modes occur where radiation can diffuse across the atmosphere faster than the sound crossing time in the gas, a reflection of weak coupling --- the relevant condition is $M_0 > \beta^{1/2}$, where $\beta \ll 1 $ is the ratio of gas pressure to radiation pressure.  Both modes depend for their existence on the presence of a magnetic field stiff enough to enforce approximately one-dimensional motion.   

The purpose of this paper is to study the nonlinear development of the buoyancy-driven waves discovered by Arons (1992), and to assess their possible effects on radiation-dominated atmospheres.  As in our earlier analysis of the nonlinear development of Gammie modes, we will use analytic techniques based on the assumption of periodic wave trains.  In keeping with the essential physical characteristics of the Arons modes, we will ignore gas pressure in our analysis.  This eliminates the possibility of obtaining a series of shock fronts, which prove to be the dominant feature of nonlinear Gammie waves. Instead, we will seek --- and find --- continuous periodic solutions.  Because the waves are driven by buoyancy (and in contrast to our earlier analysis), we must carefully account for the vertical structure of the atmosphere.  We do this by expanding our solution in powers of the vertical coordinate $z$, in addition to using a similarity variable to describe the waves.  By focusing on wavelengths that are much shorter than the scale height of the atmosphere, we are able to perform a multi-scale perturbation analysis in the radiation pressure (analogous to a WKB approximation, but nonlinear) to obtain a model for the nonlinear Arons waves. 

The plan of the paper is as follows.  In \S~2 we describe the basic equations and approximations relevant to steady-state, nonlinear waves driven by buoyancy in a radiation-dominated atmosphere.  The section concludes with a derivation of the basic equation describing the nonlinear waves.  We solve this equation in \S~3, where we also discuss the main features of the waves. In \S~4 we interpret our results and discuss the waves' global effects on atmospheres.  We emphasize that the existence of steady-state modes requires the presence of a distributed radiation source throughout the atmosphere.  Absent this source, the atmosphere is expected to collapse due to the increasing leakage of radiation with height.  We attribute the collapse of the atmospheres in the Hsu et al. (1997) simulations to this effect, and speculate that atmospheres subject to buoyancy-driven photon bubble instability will generally lose pressure support and collapse on a timescale short compared to the radiation diffusion time in the equivalent homogeneous atmosphere.  We summarize our results and present our conclusions in \S~5.

\section{Equations}

We base our model on a simplified set of radiation-hydrodynamical equations, in which the magnetic field is stiff and vertical.  The latter was assumed in Arons (1992), although Hsu et al. (1997) performed simulations for an inclined magnetic field.  All results below can be readily generalized to an arbitrary (uniform) magnetic field direction.  We assume that all vectors lie in the $x-z$ plane, and that there is a uniform gravitational field $- g \hat z$. 
 
The five equations that must be satisfied describe continuity, radiative diffusion in both the $x$ and $z$ directions, momentum conservation parallel to ${\bf B}$, and conservation of radiation energy.  Eliminating the radiation flux through the diffusion equation ${\bf F} = - (c/\kappa) \nabla p / \rho $, where $\kappa$ is the opacity (assumed constant) and $p$ is the radiation pressure, we obtain (for a vertical magnetic field)  
\begin{equation}
\label{cont}
\rho_{\tilde t} +  (\rho v)_{\tilde z} = 0 ,
\end{equation} 
\begin{equation}
\label{mom}
v_{\tilde t} + v v_{\tilde z} = - {p_{\tilde z} \over \rho} - g  ,
\end{equation} 
\begin{equation}
\label{en}
3 p_{\tilde t} + 3 v p_{\tilde z} + 4 p v_{\tilde z} = {c\over \kappa} \left[  \left({p_{\tilde z} \over \rho}\right)_{\tilde z} + \left({p_{\tilde x} \over \rho}\right)_{\tilde x}  \right] - S_0 ,
\end{equation} 
where subscripts denote partial differentiation and dimensional coordinates are surmounted by a tilde.  The source term $-S_0$ in the energy equation  represents an assumed divergence $S_0 = \nabla\cdot {\bf F_0}$ of radiation flux that is intrinsic to the fluid.  As we will show in \S~3, this added energy flux is necessary in order to obtain steady-state wave trains --- otherwise the atmosphere collapses and no steady solution is possible.  For simplicity, we will assume that $S_0$ is a constant, corresponding to a linear variation of injected radiation flux with height. This is an arbitrary choice; steady wave train solutions may exist for various dependences of $S_0$ on local state variables.   

We nondimensionalize the equations by normalizing the pressure, density, and velocity to fiducial values: $p \equiv p_0 \nu$, $\rho \equiv \rho_0 \eta$, $v \equiv v_p u$. We then define a pressure scale height $H = p_0/(\rho_0 g) \equiv  c_0^2 / g$, where $c_0$ is the isothermal sound speed associated with the radiation pressure.  Dimensionless coordinates are given by $x \equiv \tilde x /H$, $z \equiv \tilde z / H$, $t \equiv v_p \tilde t / H$, and a  characteristic Mach number is defined by $m_p \equiv v_p / c_0$.  The diffusivity parameter from Arons (1992) is given by
\begin{equation}
\label{mzero}
M_0 \equiv {c  \over \kappa \rho_0 H c_0}. 
\end{equation}
Finally, we define a dimensionless radiation source term by  $\hat \sigma \equiv H S_0/(4 m_p p_0 c_0)$. The dimensionless equations are then
\begin{equation}
\label{contnon}
\eta_{t} +  (\eta u)_{z} = 0 ,
\end{equation} 
\begin{equation}
\label{momnon}
m_p^2 (u_{t} + u u_{z} ) = - \left( 1 + {\nu_{z} \over \eta} \right) ,
\end{equation} 
\begin{equation}
\label{ennon}
3 \nu_{t} + 3 u \nu_{z} + 4 \nu u_{z} = {M_0\over m_p} \left[  \left({\nu_{z} \over \eta}\right)_{z} + \left({\nu_{x} \over \eta}\right)_{x}  \right] - 4\hat \sigma  .
\end{equation} 

We seek periodic ``plane-wave" solutions with wavelength $\lambda$, in which quantities associated with the wave depend on position and time through the combination
\begin{equation}
\label{sdef}
s = x \tan \theta + z + t ; 
\end{equation}
we henceforth denote differentiation with respect to $s$ by a prime.  Thus, the wavevector makes an angle $\theta$ with respect to the $z-$axis, as in the analysis by Arons (1992). The quantity $-v_p$ is then the $z-$component of the phase velocity. The speed of the intersection of a wave front with the $x-$axis is given by $-v_p \cot\theta$, while the overall phase speed of the waves is $v_p \cos \theta$.  

In addition to the periodic dependence on $s$ there must be a secular dependence on $z$. We will seek solutions in which the wavelength is short compared to the scale height, hence we can expand $\eta$, $\nu$ and $u$ as a Taylor series in $z$, with $|z| \ll 1$:
\begin{equation}
\label{Taylor}
\eta = \eta_0 (s) + z \eta_1 (s) + {z^2 \over 2} \eta_2 (s) + \dots \ ,
\end{equation}
with similar relations for $\nu$ and $u$.  

Since we demand that all functions of $s$ be periodic, we can define $z$-independent wave-averaged quantities by the integral 
\begin{equation}
\label{meandens}
\langle A \rangle \equiv \lambda^{-1} \int^{s+\lambda}_s A(s)  ds ;
\end{equation}
we are free to choose the density normalization so that $\langle \eta_0 \rangle = 1$.  We seek steady-state solutions with no secular mass flow, implying that 
\begin{equation}
\label{meanmassflux}
\lim_{n\rightarrow \infty} {1\over n\lambda} \int^{n\lambda}_0  \eta u  (ads + bdz)  = 0  ,
\end{equation}
for all $a$ and $b$, where 
\begin{equation}
\label{massflux}
\eta u  =  \eta_0 u_0 + z (\eta_1 u_0 + \eta_0 u_1) + {z^2\over 2} (\eta_2 u_0 + 2 \eta_1 u_1 + \eta_0 u_2) + \dots  
\end{equation}
This implies $\langle \eta_0 u_0 \rangle = 0 $ and $\eta_1 u_0 + \eta_0 u_1 = \eta_2 u_0 + 2 \eta_1 u_1 + \eta_0 u_2 =0 $.  Similarly, all higher order terms in the Taylor expansion for $\eta u$ must vanish.

Let us now consider the continuity equation.  To $O(z^0)$ we have
\begin{equation}
\label{contzero}
\eta_0' + (\eta_0 u_0)' + (\eta_1 u_0 + \eta_0 u_1) = \eta_0' + (\eta_0 u_0)' = 0 ,
\end{equation}
which is easily integrated to yield 
\begin{equation}
\label{uzero}
u_0 = {1 - \eta_0 \over \eta_0} .
\end{equation}
To $O(z^1)$ we have
\begin{equation}
\label{contone}
\eta_1' + (\eta_1 u_0 + \eta_0 u_1)' + (\eta_2 u_0 + 2 \eta_1 u_1 + \eta_0 u_2) = \eta_1' = 0 ,
\end{equation}
implying $\eta_1 = -\alpha =$ const., and 
\begin{equation}
\label{uone}
u_1 = - {\eta_1 u_0 \over \eta_0} = \alpha {1 - \eta_0 \over \eta_0^2 } .
\end{equation}
Note that the density scale height is $H/\alpha$, with  $\alpha =1$ corresponding to the $p \propto \rho \propto e^{-z}$ atmosphere considered by Arons (1992).  It is straightforward to carry the analysis of the continuity equation to higher orders in $z$ with, for example,  $\eta_2= \beta =$ const. and $u_2 = (2 \alpha^2 - \beta \eta_0)(1 - \eta_0)/\eta_0^3 $. However, these higher-order terms are not required for our analysis. 

We next consider the energy equation.  To $O(z^0)$ we have
\begin{equation}
\label{enzero}
3 (1 + u_0) \nu_0' + 3 u_0 \nu_1 + 4 \nu_0 (u_0' + u_1) =   {M_0 \over m_p } \left[  \left( {\nu_z \over \eta }\right)_0' + \left( {\nu_z \over \eta }\right)_1 + \tan^2\theta \left( {\nu_0' \over \eta_0 }\right) '\right] - 4\hat \sigma,
\end{equation} 
while to $O(z^1)$ the equation is
\begin{eqnarray}
\label{enone}
3 (1 + u_0) \nu_1' + 3 u_1 (\nu_0' + \nu_1) + 3 u_0 \nu_2 + 4 \nu_0 (u_1' + u_2) + 4 \nu_1 (u_0' + u_1) = \qquad\qquad\qquad  \nonumber \\ 
 {M_0 \over m_p } \left[  \left( {\nu_z \over \eta }\right)_1' + \left( {\nu_z \over \eta }\right)_2 + \tan^2\theta \left( {\nu_1' \over \eta_0 } - {\eta_1 \nu_0'\over \eta_0^2 }\right) '\right] , \qquad
\end{eqnarray} 
where the terms in the expansion 
\begin{equation}
\label{etanuexp}
{\nu_z \over \eta} = \left( {\nu_z \over \eta }\right)_0 +  z \left( {\nu_z \over \eta }\right)_1 + {z^2\over 2}\left( {\nu_z \over \eta }\right)_2 + \dots 
\end{equation}
are obtained from the momentum equation (\ref{momnon}):
\begin{equation}
\label{etanuzero}
\left( {\nu_z \over \eta }\right)_0 =  -1 - m_p^2 \left[ (1 + u_0)u_0' + u_0 u_1  \right] 
\end{equation}
\begin{equation}
\label{etanuone}
\left( {\nu_z \over \eta }\right)_1 =  - m_p^2 \left[ (1 + u_0)u_1' + u_0 u_2 + u_1 (u_0' + u_1) \right] . 
\end{equation}
We argue below that $(\nu_z / \eta)_2$ is not needed at the desired level of approximation.  Multiplying eq.~(\ref{momnon}) by $\eta$, we also obtain the useful results:
\begin{equation}
\label{etazero}
\nu_0' + \nu_1 =  -\eta_0 - m_p^2 \left[ u_0' + (1 - \eta_0) u_1  \right] 
\end{equation}
\begin{equation}
\label{etaone}
\nu_1' + \nu_2 =  -\eta_1 - m_p^2 \left[ u_1' + \eta_1 u_0' + \eta_0 u_0 u_2  \right] .
\end{equation}

Our expansion in $z$ is valid only if the wavelength projected on the $z-$axis is short compared to the scale height, $\tilde\lambda \ll H \cos\theta$.  (Note that this approximation just breaks down for the fastest growing linear waves, according to Arons [1992].) In this limit, the inertial terms of the momentum equation  --- those on the left-hand side of eq.~(\ref{momnon})--- should be small compared to the terms on the right-hand side. This ordering goes along with the assumption that the phase speed of the waves projected along the $z-$axis should be very subsonic, i.e., $|m_p| \ll 1$.  In this spirit, we introduce $m_p$ as a second expansion parameter for the dimensionless pressure, so that
\begin{equation}
\label{nuexp}
\nu = \nu_0 + z \nu_1 + \dots = (\nu_{00} + m_p\nu_{01} + m_p^2 \nu_{02} + \dots ) + z(\nu_{10} + m_p \nu_{11} + \dots ) + \dots
\end{equation}
We are free to set $\nu_{00} = 1$, $\nu_{10} = -1$, in order to mimic an exponential atmosphere with $p = p_0 \exp (- z/H)$, to lowest order.  By renormalizing $s$ to 
\begin{equation}
\label{wdef}
\hat w \equiv  m_p^{-1} s 
\end{equation}
(where a prime henceforth denotes differentiation with respect to $\hat w$), and treating all orders of $u$ and $\eta$ and their derivatives with respect to $w$ as $O(1)$, we can expand equations (\ref{enzero}) and (\ref{enone}) in powers of $m_p $.  In these expansions we also treat $M_0/m_p$ and $\hat \sigma$ as $\sim O(1)$.  The leading terms in both equations are then $\sim O(m_p^{-1})$; in order to describe periodic wave trains we will need the $O(m_p^{-1})$ and $O(1)$ solutions to eq.~(\ref{enzero}), and the $O(m_p^{-1})$ solution to eq.~(\ref{enone}).
 
Carrying out this program, we first note that the terms proportional to $(\nu_z/\eta)_1$ in eq.~(\ref{enzero}) and $(\nu_z/\eta)_2$ in eq.~(\ref{enone}) are $\sim O(m_p)$, so they can be ignored. Next, by expanding equations (\ref{etazero}) and (\ref{etaone}) we obtain:
\begin{equation}
\label{nuexp2}
\nu_{01}' = 1 - \eta_0; \qquad\qquad \nu_{02}' = - (\nu_{11} + u_0'); \qquad\qquad \nu_{11}' = - (\eta_1 + \nu_{20}) . 
\end{equation}
These results allow us to write down eq.~(\ref{enzero}) to $O(m_p^{-1})$:
\begin{equation}
\label{enzeromp0}
4 u_0' = {M_0\over m_p} \tan^2\theta \left(  {\nu_{01}'\over \eta_0}\right)' = {M_0\over m_p} \tan^2\theta \left( - {\eta_0'\over \eta_0^2}\right) = {M_0\over m_p} \tan^2\theta \ u_0' 
\end{equation}
(where we have used eq.~[\ref{uzero}]), implying
\begin{equation}
\label{enzeromp1}
m_p = {1\over 4} M_0 \tan^2\theta  . 
\end{equation}
Now consider eq.~(\ref{enone}) to $O(m_p^{-1})$:
\begin{equation}
\label{enonemp1}
4 (u_1' - u_0') = {M_0\over m_p} \tan^2\theta \left(  {\nu_{11}'\over \eta_0} - {\eta_1\nu_{01}'\over \eta_0^2}\right)'  .
\end{equation}
Substituting $\eta_1 = -\alpha$ and using equations (\ref{uzero}), (\ref{uone}), and (\ref{enzeromp1}), we obtain $\nu_{11}' = \eta_{0} - 1 = - \nu_{01}'$, where we have imposed the condition $\langle \nu_{11}' \rangle = 0$ for a periodic solution. We may then set $\nu_{11} = - \nu_{01}$.

Finally, we consider the $O(1)$ terms in eq.~(\ref{enzero}). Using previous results, we find that $3 (1+ u_0) \nu_{01}' + 3 u_0 \nu_{10} = 0$.  Therefore, the surviving terms are
\begin{equation}
\label{enzerozero}
4 \nu_{01}u_0' + 4 u_1 = {M_0\over m_p} \left[ \left( {\eta_0'\over \eta_0^3}\right)' + \tan^2\theta  \left( {\nu_{02}'\over \eta_0}\right)' \right] - 4\hat \sigma = {M_0\over m_p} \left[ \left( {\eta_0'\over \eta_0^3}\right)' + \tan^2\theta  \left( {\nu_{01} - u_0' \over \eta_0}\right)' \right] - 4\hat \sigma ,
\end{equation}
where we have used eq.~(\ref{nuexp2}) and substituted  $\nu_{11} = - \nu_{01}$.  Using the result that $\nu_{01}'/\eta_0 = u_0$ and expressing $u_0$ and $u_1$ in terms of $\eta_0$ using equations (\ref{uzero}) and (\ref{uone}), we obtain
\begin{equation}
\label{etaeq1}
\left( {\eta_0'\over \eta_0^3}\right)' =  \sin^2\theta {(1- \eta_0)(\alpha - \eta_0 ) \over \eta_0^2}   - \hat \sigma ,
\end{equation}
which is the equation describing steady wave trains. To simplify our study of solutions, we renormalize the independent variable to $w = \hat w \sin\theta $ and set $\sigma \equiv \hat\sigma / \sin^2\theta$, thus eliminating the factor $\sin^2\theta$ from eq.~(\ref{etaeq1}). The basic equation for our study of wave solutions is therefore
\begin{equation}
\label{etaeq}
\left( {\eta'\over \eta^3}\right)' =  {(1- \eta)(\alpha - \eta) \over \eta^2}   - \sigma ,
\end{equation}
where a prime now denotes differentiation with respect to $w$ and we have dropped the subscript ``0" from $\eta_0$.

\section{Solutions}

We seek continuous periodic solutions of eq.~(\ref{etaeq}), subject to the constraint that $\langle \eta \rangle = 1$.  That such solutions might exist is evident from the fact that eq.~(\ref{etaeq}) can be written in the form
\begin{equation}
\label{genericODE}
{1\over 2} {d\over dy} \left( y'\right)^2 + {df(y) \over dy}  = 0
\end{equation}
where $y$ is a function of $\eta$ and $y' = \eta' dy/d\eta$.  $f(y)$ is also an implicit function of $\eta$ through $y$. This is, of course, just a generalized version of the Hamiltonian equation of energy conservation in a potential.  In order for eq.~(\ref{genericODE}) to have periodic solutions, $f(\eta)$ must have two real roots, $f(\eta_+) = f(\eta_-) = 0$, with $f < 0$ for $\eta_- < \eta < \eta_+$,  and the integral
\begin{equation}
\label{genericint}
\int^{\eta_+}_{\eta_-} {dy\over d\eta} {d\eta\over  \left( - 2f \right)^{1/2}} , 
\end{equation}
which equals half the wavelength, must be finite.

The requirement of an intrinsic energy source in the atmosphere, $\sigma \neq 0$, is also apparent from the form of eq.~(\ref{etaeq}).  If we choose $\alpha = 1$, for example, then the first term on the right-hand side of eq.~(\ref{etaeq}) is $\geq 0$, precluding a periodic solution unless $\sigma > 0$.  In general, for a given $\alpha$ and wavelength, $\sigma$ is an eigenvalue of the problem. 

Equation (\ref{etaeq}) is most easily solved by choosing a maximum density, $\eta_+$, then integrating to find the wavelength, minimum density $\eta_-$, and $\sigma$, subject to the mean density constraint.  It is better to start integrating from the maximum density rather than the minimum density because, in solutions with large density contrasts, $\eta_+$ is exponentially sensitive to $\eta_-$. 

Figure 1 shows the density profile of the waves for $\eta_+ =$ 1.1, 2, and 11, with $\alpha = 1$. Parameters for a wider range of solutions are shown in Table 1.  Even for wave trains with very low amplitudes, $\eta_+ - 1 \ll 1$ (Fig.~1a), the density pattern is far from sinusoidal.  At low amplitudes, $\eta_+ - 1 \la 1$, the depression in the low-density portion of the wave is sharper and deeper than the enhancement in the dense region. This behavior switches markedly for $\eta_+ - 1 \ga 1$, and for high density contrasts (Fig. 1b) the dense regions are extremely sharp and narrow, while the low density regions are broad and flat.  

Figure 2 and Table 1 illustrate the variation of minimum density $\eta_-$, wavelength $\lambda$ (in units of $w$), and intrinsic flux divergence $\sigma$ with $\log_{10}(\eta_+ - 1)$.  The minimum density varies in the opposite sense to the maximum density, but at large density contrasts this variation is slower than logarithmic.  The wavelength decreases with increasing density contrast, also slower than logarithmically, while the intrinsic radiation source $\sigma$, required for a steady-state solution, increases slightly faster than logarithmically, although still very slowly.   

Interestingly, the inverse relationship between density contrast and wavelength is opposite to that found in the nonlinear shock train solutions (Begelman 2001).  Numerical simulations (Turner et al.~2005) indicate that the shock train solutions evolve toward longer wavelengths, higher density contrasts, and larger luminosities.  If higher density contrasts and larger luminosities are also energetically favored in the case of buoyancy-driven photon bubbles, then we would expect evolution toward {\it shorter} wavelengths.  Indeed, the fastest-growing Arons modes have long wavelengths, close to the pressure scale height (see \S~4.2), whereas the fastest-growing Gammie modes are short.

\begin{table*}
\begin{tabular}{ccccccccccccc}
\small
 \ \ &\vline&&$\alpha = 1$&&\vline&&$\alpha=0.5$&&\vline&&$\alpha=2$\\
\hline
$\log_{10}(\eta_+ - 1)$&
\vline&$\eta_-$&$\lambda[w]$&$\sigma$&
\vline&$\eta_-$&$\lambda[w]$&$\sigma$&
\vline&$\eta_-$&$\lambda[w]$&$\sigma$\\
 \hline
-2&\vline&0.979&82.2&0.000111&\vline&---&---&---&\vline&0.990&6.29&0.000146\\
-1&\vline&0.839&25.9&0.00979&\vline&---&---&---&\vline&0.909&6.24&0.0136\\
0&\vline&0.478&9.44&0.395&\vline&0.308&17.2&0.386&\vline&0.547&5.18&0.722\\
1&\vline&0.270&6.11&2.94&\vline&0.225&10.1&1.49&\vline&0.287&3.97&6.16\\
2&\vline&0.186&5.39&8.30&\vline&0.170&8.44&3.79&\vline&0.193&3.62&17.6\\
3&\vline&0.143&5.11&16.2&\vline&0.134&7.79&7.43&\vline&0.146&3.49&34.1\\
4&\vline&0.116&4.96&26.55&\vline&0.111&7.44&12.2&\vline&0.118&3.41&55.5\\
\end{tabular}
\caption{Parameters characterizing periodic solutions of the wave train equation (\ref{etaeq}), for three values of the density gradient parameter, $\alpha$. All solutions satisfy the mean density constraint $\langle \eta \rangle = 1$. The wavelength $\lambda$ is in units of the dimensionless variable $w$. Solutions do not exist for $\alpha = 0.5$ and $\eta_+ = 1.01, 1.1$ --- see \S~3.1 for an explanation.}
\label{tbl1}
\end{table*}
\begin{figure*}
\plotone{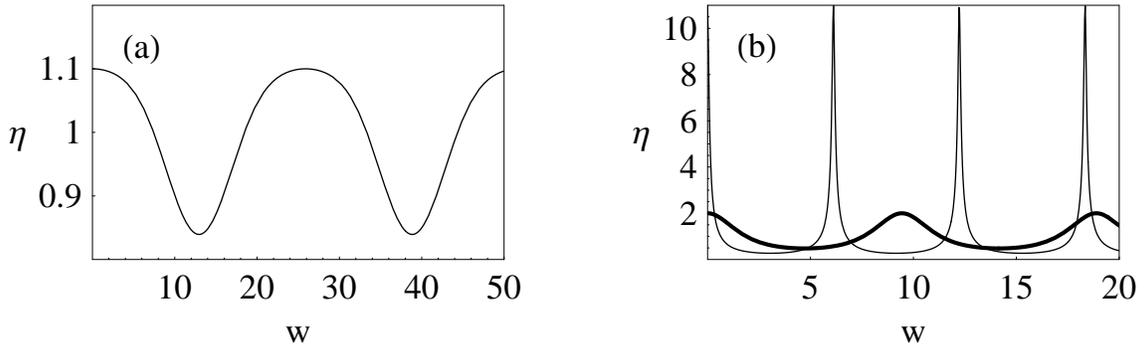}
\caption{Numerical solutions of the wave train equation (\ref{etaeq}) with $\alpha=1$, subject to the mean density constraint. Parameters for the plotted solutions can be found in the second, third and fourth rows of Table 1. (a) Solution with weak density contrast, $\eta_+ = 1.1$.  (b) Superposed solutions with moderate ($\eta_+ = 2$; heavy line) and large ($\eta_+ = 11$; light line) density contrasts. } 
\label{fig1}
\end{figure*}

\begin{figure*}
\plottwo{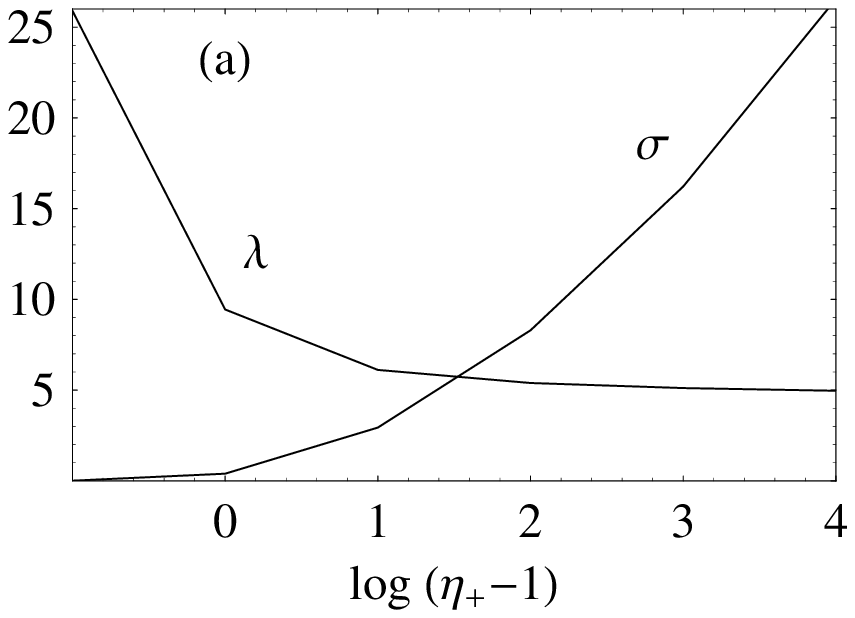}{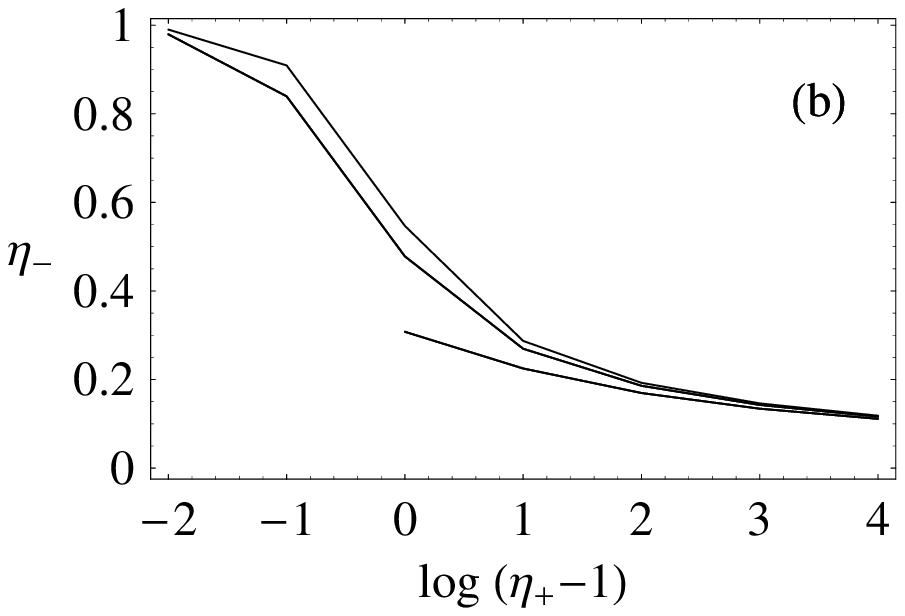}
\caption{(a) Plots of $\sigma$ and $\lambda$ vs. $\log_{10}(\eta_+ -1)$, for solutions with $\alpha = 1$.  (b) Minimum density $\eta_-$ vs.  $\log_{10}(\eta_+ -1)$ for $\alpha= 2, 1, 0.5$ (upper to lower curve). }
\label{fig2}
\end{figure*}

\subsection{Dependence on Density Gradient}

The parameter $\alpha$ represents the underlying density gradient in the atmosphere, with $\alpha = 1$ corresponding to the ``isothermal" atmosphere $p\propto\rho$ assumed by Arons (1992). The radiation entropy gradient $\nabla \ln (p/\rho^{4/3}) \sim 4\alpha/3 - 1$, so all atmospheres with $\alpha > 0.75$ are convectively stable according to the Schwarzschild criterion.  Note that, because $M_0 < 1 $ for atmospheres subject to buoyancy-driven photon bubbles, it may be  possible for large convective cells to transport radiation energy without excessive leakage. (Recall that $M_0$ is the ratio of the dynamical time to the diffusion time across a scale height.) The models with $\alpha = 1$ and $\alpha =2$ correspond to convectively stable atmospheres, while the model with $\alpha = 0.5$ is strongly unstable to convection. 

From Fig. 2b, it is apparent that density contrasts are relatively insensitive to $\alpha$.  The one glaring exception to this statement is the fact that no periodic solutions exist for low density contrasts, in the case of $\alpha = 0.5$. This is easily understood by considering the right-hand side of eq.~(\ref{etaeq}).  In order to have a periodic solution, the right-hand side must be negative at the upper turning point ($\eta_+$) and positive at the lower turning point ($\eta_-$).  For sufficiently weak density contrasts and $\alpha < 1$, it is possible to have $\alpha < \eta_- < 1$.  In this case, the first term on the right-hand side is negative, implying that $\sigma$ must be negative in order to make the right-hand side positive.  But at $\eta_+$ the first term is positive implying that $\sigma $ must be positive, leading to a contradiction. 

The quantities $\sigma$ and $\lambda$ are more sensitive to the density gradient than is $\eta_-$. At fixed $\eta_+$, $\sigma$ increases and $\lambda$ decreases with $\alpha$.  Since larger $\alpha$ means a steeper density gradient, these trends might reflect the increasing effectiveness of buoyancy in driving radiation through the atmosphere via the low-density channels.  Larger $\sigma$ means that more energy is being transported by diffusion through the photon bubbles, while a smaller wavelength could mean that a smaller fraction of a pressure scale is sufficient to get the required driving force. Note that this is opposite to the trend for ordinary convection, in which energy is assumed to be advected by fluid motions, and energy transport is expected to be larger for smaller $\alpha$.  In buoyancy-driven photon bubble instability, diffusion rather than advection transports the energy.  

\section{Interpretation}

\subsection{Physical Properties of Solutions}

The dimensional wavelength of solutions, $\tilde \lambda$, is related to the dimensionless wavelength $\lambda$ (in units of $w$), by
\begin{equation}
\label{lambdatilde}
\tilde \lambda = m_p H  \cot\theta \  \lambda . 
\end{equation}
Combining this equation with eq.~(\ref{enzeromp1}), we obtain
\begin{equation}
\label{kh}
{\tilde\lambda \over H } = {\lambda\over 4} M_0 \tan\theta  . 
\end{equation}
From our numerical solutions (Table 1) we have determined that $ \lambda /4 \sim O(1)$ for the cases with high density contrast.   In order for our analysis to be valid, however, it is not sufficient for the wavelength merely to be smaller than the pressure scale height.  A stronger condition  --- required by our expansion in powers of $z$ --- is that the wavelength projected onto the $z-$axis, $\tilde\lambda_z = \tilde\lambda / \cos\theta$, must be smaller than $H$. Thus, a necessary condition for the validity of our analysis is that 
\begin{equation}
\label{kzh}
M_0 < {\cos^2\theta \over \sin\theta }  . 
\end{equation}
From eq.~(\ref{enzeromp1}) this condition implies that $m_p < \sin\theta / 4 < 1$, thus ensuring the validity of our expansion in $m_p$.

In all cases $\sigma$ is positive, implying that the atmosphere itself must supply extra energy at all heights, in order to support a steady-state wave solution. In physical units, the intrinsic flux radiated by the atmosphere is 
\begin{equation}
\label{Szero}
F_0 \sim S_0 H = 4 \sigma \sin^2 \theta \ m_p p_0 c_0
\end{equation}

Since $\sigma \ga O(10)$ for the high contrast solutions, these atmospheres must radiate furiously.  By definition, the energy loss rate due to diffusion in the equivalent homogeneous atmosphere is $\sim M_0 p_0 c_0$.  The inhomogeneous atmospheres in our model are losing energy $\sigma \sin^2 \theta \tan^2\theta $  times faster --- possibly leaking energy on less than a dynamical time.

It is important to remember that our adoption of an intrinsic flux divergence $S_0$ was arbitrary --- we took this step only because a steady-state wave solution could not be obtained without it. What would happen if, as is more likely, such a powerful intrinsic source term is absent?  Although this case is not covered explicitly by our calculations, the only sensible deduction would seem to be that the atmosphere loses radiation pressure support and collapses on the leakage time or the dynamical time, whichever is longer.  Indeed, since the flux used in our radiation hydrodynamical equations is defined relative to the fluid frame, one way for the atmosphere to generate a flux divergence is for the gas to sink in the gravitational potential.  We precluded this possibility by demanding $\langle \rho v \rangle = 0$, for mathematical reasons, but this outcome cannot be excluded physically.  

\subsection{Relationship to Arons Instability}

While we have not proven that our nonlinear wave trains represent the outgrowth of Arons's (1992) instability,  there are a number of aspects which make this identification plausible.  First, our wave trains are driven by buoyancy forces --- we guaranteed this by setting the gas pressure to zero (as did Arons in his linear analysis) and treating inertial terms as higher order than buoyancy terms.  Second, we exploited the same near-cancellation in the flux-divergence as Arons did.  In his case, a slight imbalance drove the instability by diffusing more energy into the lower parts of bubbles than is removed at the top. In our case, it supplied the crucial driving term $(\eta'/\eta^3)'$ in the wave equation (\ref{etaeq}).  Third, our waves only appear in the $O(z)$ terms of an expansion in height, further illustrating that they are driven by buoyancy effects. Fourth, the vertical component of the wave phase velocity is negative, in agreement with the linear instability (Arons 1992, equations [19] and [42]; in contrast to the group velocity of the linear waves, which is positive --- see Arons eq.~[22]).  Finally, the wave speeds in our model scale $\propto M_0$, indicating that the wave speed is set by the diffusion of radiation across the bubble, as in the linear instability. 
 
According to the linear analysis, the most rapidly growing modes have $kH \cos\theta = 0.66$ (Arons 1992, eq.~[44]), where $k$ is the wavenumber. Projected on the $z-$axis, the wavelength is about 10 times longer than the scale height, putting it outside the limit of validity for our analysis.  The fastest growing linear mode that is (marginally) consistent with our analysis has $\tilde \lambda_z \sim H$. If we assume that this mode determines the orientation of the nonlinear wave pattern, then the inequality in eq.~(\ref{kzh}) is replaced by an approximate equality. In the limit $M_0 \ll 1$, we then have $\cos\theta \sim \sqrt M_0 \ll 1$, and the dominant density contours will tend to align with the magnetic field.  Thus, our nonlinear analysis reproduces another key feature of the linear instability, as well as the numerical simulations by Hsu et al.~(1997).  

Finally, we note that the only existing numerical simulations of buoyancy-driven photon bubbles (Hsu et al. 1997), while reproducing the linear growth rates adequately, are plagued by the rapid collapse of the simulated atmosphere as the waves saturate. We conjecture that this outcome is real and reflects the lack of a compensating intrinsic source of radiation in the simulations.  The fastest growing modes grow on a dynamical timescale (Arons 1992, eq.~[45]), which is similar to the collapse time.   

\subsection{Minimum Magnetic Field Strength and Effects of Gas Pressure}

The radiation pressure perturbations associated with the waves is $\Delta p \sim  (\tilde\lambda / H) p_0 \tan\theta \sim M_0 p_0 \tan\theta $.  These pressure fluctuations will be largely compensated by fluctuations in the magnetic field strength, in order to maintain overall pressure balance in the $x-$direction. The change in the magnetic pressure $p_B$ is related to sideways compressional or expansional motion by $\Delta p_B \sim (\Delta \tilde x / \tilde \lambda) p_B$, where $\Delta \tilde x$ is the sideways displacement caused by the passage of the wave.  Now, in order for the magnetic field to enforce approximately one-dimensional motion, we require that the sideways displacement be much smaller than the parallel displacement of a typical fluid element, $\Delta \tilde x \ll \tilde\lambda / \cos\theta$.  This implies
\begin{equation}
\label{maglimit}
{p_B \over p_0} \gg {\tilde \lambda \over H} \sin\theta = {\lambda\over 4} M_0 {\sin^2\theta \over \cos\theta}.
\end{equation}
Equation (\ref{maglimit}) appears somewhat different from the analogous criterion presented by Arons (1992) in connection with the linear instability. It is beyond the scope of this paper to carry out a fully magnetohydrodynamical analysis.

The effects of gas pressure will be most important in the densest regions of the waves, where $\eta \approx \eta_+$.  Rapid energy exchange between matter and radiation will enforce approximately isothermal conditions across the wavefronts, implying that the gas pressure is proportional to the local density. If the mean gas pressure is $\beta_g p_0$, where $\beta_g \ll 1$, then gas pressure will affect the wave dynamics when $\beta_g \eta_+ p_0$ approaches $\Delta p$, or equivalently, when 
\begin{equation}
\label{gaslimit}
\eta_+ \sim \beta_g ^{-1} {\tilde \lambda \over H} \tan\theta .
\end{equation}
In such cases, gas pressure may limit the maximum density but is unlikely to have much effect on the values of $\lambda$ or $\sigma$.

\section{Discussion and Conclusions}

We have produced an analytic model for periodic, nonlinear waves driven by buoyancy in a radiation pressure-dominated atmosphere with a stiff magnetic field.  To isolate the effects of buoyancy, we ignored gas pressure and performed a multi-scale expansion in powers of the vertical coordinate $z$ and wave Mach number $m_p$. All solutions for steady-state, periodic wave structures require the existence of a large radiation source term in the atmosphere, which we modeled as a uniform, positive flux divergence term in the energy equation.  Physically, we believe that this source term compensates for the enhanced loss of radiation via the waves.  Without it the atmosphere would lose pressure support and collapse, on a timescale much shorter than the diffusion time associated with the equivalent homogeneous atmosphere. 

Based on several similarities of generic properties and physical mechanisms, we identify our wave solutions with the nonlinear development of the photon bubble instability discovered by Arons (1992). Simulations of the linear growth and saturation of this instability (Hsu et al. 1997) were plagued by the rapid collapse of the simulated atmosphere shortly after the waves reached nonlinear amplitudes.  We suggest that this collapse resulted from the rapid release of energy via the waves, as indicated by our analytic model.

Buoyancy-driven photon bubbles occur primarily when the radiation diffusion timescale across a pressure scale height is longer than the dynamical time, corresponding to the parameter $M_0$ being smaller than one. (It may be possible for buoyancy-driven photon bubbles to develop in atmospheres with $M_0 \ga 1$, provided that the wavefronts are sufficiently oblique to the magnetic field.) This condition is satisfied in a number of astrophysical systems, including the equatorial regions of radiation-dominated accretion disks (where $M_0 \sim \alpha$, the viscosity parameter), accretion columns onto neutron stars in X-ray binaries, massive stellar envelopes well below the photosphere, and gamma-ray burst fireballs.  Whether the linear instability develops and saturates in any of these systems is not clear, because the growth of the instability is rather slow.  For example, in accretion disks subject to the magnetorotational instability (MRI), the magnetic field structure is expected to change on a timescale comparable to the dynamical (Keplerian) time. This is shorter than the growth time of all but (possibly) the fastest growing buoyancy-driven photon bubbles.  Hence, it would not be surprising if buoyancy-driven photon bubbles were suppressed in accretion disks. (Gas pressure-driven photon bubbles, on the other hand, can grow faster than MRI, and are expected to be important in accretion disks [Begelman 2002, 2005; Turner et al. 2005].) In accretion columns onto a neutron star polar cap, however, the magnetic field is anchored into the neutron star crust and its structure is fixed.  We would expect buoyancy-driven photon bubbles to grow to nonlinear amplitudes under these circumstances, perhaps leading to the collapse of the accretion column. 

What would happen to an atmosphere that collapsed under the action of buoyancy-driven photon bubbles? If we treat the collapsing atmosphere as a sequence of hydrostatic models, then both the diffusion parameter, $M_0$ ($\propto c_0^{-1}$ for a fixed column density), and the ratio of gas pressure to radiation pressure, $\beta_g$, are expected to increase. If $M_0$ became large, then the atmosphere would enter the strong-diffusion regime and gas pressure-driven shock trains would become the dominant form of photon bubble.  Alternatively, $\beta_g$ could increase until gas pressure halted the collapse.  A third possibility, which may be relevant in the case of neutron star accretion columns, is that the gravitational binding energy liberated by the accretion flow provides at least part of the compensating flux divergence. It is not clear, however, whether this would be adequate to prevent the collapse of the column.  

\acknowledgments
I am grateful to Jon Arons for pointing out the atmospheric collapse problem in his 1997 simulations with Hsu and Klein. This research was supported in part by NASA Astrophysical Theory Program grant NAG5-12035 and by the National Science Foundation under Grants AST-0307502 and PHY-990794.  Part of this work was carried out at the Kavli Institute for Theoretical Physics at the University of California, Santa Barbara; I thank the members of KITP for their hospitality.

\end{document}